\documentclass[conference]{IEEEtran}
\IEEEoverridecommandlockouts
\usepackage{cite}
\usepackage{amsmath,amssymb,amsfonts}
\usepackage{algorithmic}
\usepackage{graphicx}
\usepackage{textcomp}
\usepackage{xcolor}
\def\BibTeX{{\rm B\kern-.05em{\sc i\kern-.025em b}\kern-.08em
    T\kern-.1667em\lower.7ex\hbox{E}\kern-.125emX}}

\usepackage{amsthm}

\newtheorem{theorem}{Theorem}[section]
    
\newtheorem{lemma}[theorem]{Lemma}
\theoremstyle{remark}

\begin{document}

\title{Efficient Ordered-Transmission Based Distributed Detection under Data Falsification Attacks}

\author{\IEEEauthorblockN{Chen Quan$^{1}$, Nandan Sriranga$^{1}$, Haodong Yang$^{1}$, Yunghsiang S. Han$^{2}$}, \IEEEmembership{Fellow, IEEE},
Baocheng Geng$^{3}$ and Pramod K. Varshney$^{1}$, \IEEEmembership{Life Fellow, IEEE}

\IEEEauthorblockA{
\textit{$^{1}$Syracuse University\ $^{2}$University of Electronic Science and Technology of China \ $^{3}$University of Alabama at Birmingham }
}}

\maketitle

\begin{abstract}
In distributed detection systems, energy-efficient ordered transmission (EEOT) schemes are able to reduce the number of transmissions required to make a final decision. 
%This is possible through the transmission of the least amount of data, which is necessary to arrive at a decision satisfying the required performance metrics at the fusion center (FC). 
In this work, we investigate the effect of data falsification attacks on the performance of EEOT-based systems. We derive the probability of error for an EEOT-based system under attack and find an upper bound (UB) on the expected number of transmissions required to make the final decision. Moreover, we tighten this UB by solving an optimization problem via integer programming (IP). We also obtain the FC's optimal threshold which guarantees the optimal detection performance of the EEOT-based system. Numerical and simulation results indicate that it is possible to reduce transmissions while still ensuring the quality of the decision with an appropriately designed threshold.
% Numerical and simulation results show that the impact of data falsification attacks on an EEOT-based system can be reduced if the threshold used by the FC is selected appropriately.
\end{abstract}

\section{Introduction}
Energy efficiency is an important design consideration to enhance the lifetime of a wireless sensor network (WSN). Several notable schemes have been proposed in the past to improve the energy efficiency of WSNs~\cite{rago1996censoring,bandyopadhyay2003energy,bajovic2011sensor,blum2008energy}. In this paper, we consider the energy-efficient ordered transmission (EEOT) scheme first proposed in \cite{sriranga2018energy}, for a distributed WSN. In the EEOT scheme, each sensor sends its local binary decision to the fusion center (FC) at a time that is proportional to the inverse of the absolute value of the corresponding log-likelihood ratio (LLR). Consequently, more informative sensors transmit prior to less informative ones. After the FC has received enough observations to make a final decision of desired quality, it broadcasts a stop signal to prevent further transmissions from sensors that have not yet transmitted.

The ordered transmission (OT) scheme was first proposed in \cite{blum2008energy}, where only informative sensors\footnote{Sensors with larger LLR magnitudes are more informative than the ones with smaller LLR magnitudes. In the rest of the paper, the sensors with larger LLR magnitudes are referred to as informative sensors. } report their LLRs to the FC, rather than all the sensors transmitting raw data.
In \cite{braca2011single}, the authors have demonstrated that a single observation is sufficient to make a final decision for an ordered transmission based (OT-based) system with a large number of sensors. The authors of \cite{sriranga2018energy} proposed an energy-efficient OT (EEOT) scheme in which informative sensors send binary decisions, rather than sending LLRs, to the FC. A correlation-aware OT scheme is proposed in \cite{gupta2020ordered2}, where spatial correlation between the sensors is considered. According to the studies in \cite{blum2008energy,braca2011single,sriranga2018energy,gupta2020ordered2}, the OT-based schemes are capable of efficiently reducing the number of transmissions needed for decision-making. 

The OT approach has also been used in some other systems to reduce communication costs and energy consumption. In \cite{rawas2011energy}, the OT scheme was used for non-coherent networked signal detection, where the LLRs at the sensors only have non-negative values available, to reduce transmissions in the network. In \cite{hesham2012distributed}, sequential detection along with OT was considered for cooperative spectrum sensing to obtain fast and reliable decisions regarding primary user activities over the spectrum. The authors of \cite{chen2020optimal} applied the OT scheme to the distributed quickest change detection problem, where the number of transmissions to the FC are reduced without affecting the detection delay of the system. In \cite{gupta2020ordered}, the OT scheme was incorporated into energy harvesting sensor networks in order to increase the energy efficiency of the sensors. The authors of \cite{chen2020ordered} presented an ordered gradient approach to eliminate some sensor-to-FC uplink communications in distributed ADMM. 

Thus, the OT scheme is a promising scheme for a significant improvement in the energy efficiency of distributed systems. However, due to the open nature of WSNs and the large-scale deployment of low-cost sensors, the performance of the EEOT-based systems in the presence of data falsification attacks is an important aspect to consider. We consider the data falsification attack model used in \cite{6582732,kailkhura2016data}, where a sensor may be malicious that sends falsified data to the FC with the intention of degrading the detection performance of the system. To the best of our knowledge, this is the first work that investigates the impact of data falsification attacks on OT-based schemes. We evaluate the performance of the EEOT-based system via its detection performance and the average number of transmissions required in the presence of data falsification attacks. We derive the probability of error for the EEOT-based distributed detection system along with the optimal decision threshold at the FC. Further, we derive a tight upper bound (UB) on the number of transmissions required under attack. Numerical and simulation results indicate that it is possible to reduce transmissions while still ensuring the quality of the decision with an appropriately designed threshold.

In the following sections of the paper, we first present the system model in Section \ref{sec:system_model_EEOT} and evaluate the performance of the EEOT-based system under data falsification attacks. The numerical results are presented in Section \ref{sec:simulation_EEOT} and the paper is concluded in Section \ref{sec:conclusion_EEOT}.

\section{System Model}
\label{sec:system_model_EEOT}
A distributed OT-based network consisting of $N$ sensors and one FC is considered in this work. A binary hypothesis testing
problem is investigated where hypothesis $\mathcal{H}_1$ indicates the presence of the signal and $\mathcal{H}_0$ indicates the absence of the signal, and the goal is to determine which of the two hypotheses is true. Let $y_i$ be the received observation at sensor $i\in\{1,2,\dots,N\}$. We assume that all the observations are independent and identically distributed (i.i.d) conditioned on the hypotheses. For sensor $i$, the observation $y_i$ is modeled as 
\begin{align}
\label{eq:obs_honest}
y_i = 	\begin{cases}
						n_i&\text{under $\mathcal{H}_0$}\\
						s+n_i&\text{under $\mathcal{H}_1$},    	
					\end{cases}
\end{align}
where $n_i$ is the Gaussian noise with zero mean and variance $\sigma^2$ and $s$ is the signal strength at each sensor. $s$ and $n_i$ are assumed to be independent. Hence, $y_i|\mathcal{H}_0$ and $y_i|\mathcal{H}_1$ follow Gaussian distributions, i. e., 
\begin{align}
    y_{i}|\mathcal{H}_0\sim\mathcal{N}(0,\sigma^2)\\
    y_{i}|\mathcal{H}_1\sim\mathcal{N}(s,\sigma^2),\label{eq:y_i_h1}
\end{align}
respectively. Based on the local observations $\{y_i\}_{i=1}^N$, each sensor $i\in\{1,\dots,N\}$ makes a binary decision $v_i\in\{0,1\}$ regarding the two hypotheses using the LRT
\begin{equation}
    L_i=\log\left(\frac{f_{Y_i}(y_i|\mathcal{H}_1)}{f_{Y_i}(y_i|\mathcal{H}_0)}\right)\overset{v_i=1}{\underset{v_i=0}{\gtrless}}\lambda,
\end{equation}
where $L_i$ denotes the LLR of sensor $i$, $\lambda=\log\left(\frac{\pi_0}{\pi_1}\right)$ is the identical threshold used by all the sensors and $v_i$ is the local decision made by sensor $i$. According to the OT-based framework presented in \cite{blum2008energy} and \cite{sriranga2018energy}, the sensor transmissions are ordered based on the magnitudes of their LLRs. If the magnitudes of the LLRs are ordered as $|L_{[1]}|>|L_{[2]}|>\ldots>|L_{[N]}|$, the sensors transmit their local decisions in the order determined by their magnitude-ordered LLRs, i.e., in the order of $v_{[1]},v_{[2]},\dots,v_{[N]}$, where $v_{[k]}$ is the local decision made by the $[k]^{th}$ sensor. 
% Hence, $v_{[k]}$ is given as
% \begin{equation}
%     L_{[k]}\overset{v_{[k]}=1}{\underset{u_{[v]}=0}{\gtrless}}\lambda.
% \end{equation}
It is important to note that the magnitude-ordered LLRs do not imply that the local decisions are also magnitude-ordered, i.e., $|L_{[1]}|>|L_{[2]}|>\ldots>|L_{[N]}|$ does not imply $v_{[1]}\geq v_{[2]}\geq \ldots\geq v_{[N]}$.

The decision rule used by the FC is the same as in \cite{sriranga2018energy} and it is given by
\begin{equation}\label{eq:EE-rule}
\left\{
\begin{array}{lcl}
    \sum_{i=1}^kv_{[k]}\geq  T&&\text{decide $\mathcal{H}_1$}\\
    \sum_{i=1}^kv_{[k]}< T-(N-k)&&\text{decide $\mathcal{H}_0$}
\end{array}\right.,
\end{equation}
where $T$ is the threshold used by the FC. As in \cite{sriranga2018energy}, the following assumption is made in this paper.

{\em Assumption 1:} $Pr(v_i=1|\mathcal{H}_1)\rightarrow{1}$ and $Pr(v_i=0|\mathcal{H}_0)\rightarrow{1}$ when $s\to\infty$.

%\begin{rem}
%Note that large $s$ and the trustworthiness of sensors are key to prove the result that the average number of transmissions saved by utilizing the EEOT-based scheme are lower bounded by $N/2$ (see %\cite{sriranga2018energy}). However, when $s$ is small or when there are Byzantine nodes in the system, Assumptions 1 is no longer valid. In this paper, we consider the case where there are Byzantines in the network.
%\end{rem}

\subsection{Attack Model}
In this sub-section, we discuss the attack model used by the malicious nodes in this paper. In our setup, a sensor $i$ can be honest $(H)$ or malicious $(M)$. Assume that each sensor in the network has $\alpha$ probability of being malicious, i.e., $P(i=M)=\alpha$. We also assume that the malicious sensors falsify data by flipping their local decisions sent to the FC. Let $p=p(v_i\neq u_i|i=M)$ denote the probability that the malicious node $i$ flips its local decisions. Hence, we have 
\begin{equation}
    \left\{
    \begin{array}{rcl}
        v_i= 1-u_i&\text{with probability $\alpha p$}\\
        v_i= u_i&\text{with probability $1-\alpha p$}
    \end{array}\right.,
\end{equation}
where $v_i$ is the original unaltered local decision made by sensor $i$ and $u_i$ is the local decision sent to the FC by sensor $i$. Note that $v_i$ is not necessarily equal to $u_i$ due to the action of the malicious nodes. In the next sub-section of this paper, we evaluate the performance of the EEOT-based system in the presence of malicious nodes in terms of detection performance and the average number of transmissions required in the network. 

\subsection{Detection Performance}
We begin our analysis of the detection performance of the EEOT-based scheme in the presence of malicious nodes by first presenting the following Lemma. It states that in the presence of malicious nodes, the EEOT-based system can achieve the same detection performance as the one without ordering.
\begin{lemma}
\label{lem:detect_perf_ot}
When the FC follows the Bayesian decision rule, the detection performance of systems with and without the use of the EEOT-based scheme are the same in the presence of data falsification attacks. 
\end{lemma}
\begin{IEEEproof}
Please see Appendix~\ref{Appendix:EEOT}.
\end{IEEEproof}

Thus, we evaluate the detection performance of the EEOT-based system in the presence of data falsification attacks by evaluating the detection performance of the corresponding distributed system without ordering. According to \cite{quan2022enhanced}, for the system without ordering, the probabilities of $v_i=1$ and $v_i=0$ given $\mathcal{H}_h$ are expressed as
\begin{align}\label{eq:u_i_H}
    \pi_{1,h}=P(v_i=1|\mathcal{H}_h)=Q\left(\frac{\lambda-\mu_h}{\nu_h}\right)
\end{align}
and $\pi_{0,h}=P(v_i=0|\mathcal{H}_h)=1-\pi_{1,h}$, respectively, for $h=0,1$, where $Q(.)$ is the tail distribution function of the standard normal distribution, $\mu_0=0$, $\mu_1=s$ and $\nu_0=\nu_1=\sigma$. Hence, the probabilities of $u_i=1$ and $u_i=0$ given $\mathcal{H}_h$ are expressed as
\begin{align}\label{eq:u_i}
    \widetilde{\pi}_{1,h}=&P(u_i=1|\mathcal{H}_h)\notag\\
    =&P(u_i=1|\mathcal{H}_h,i=M)P(i=M)\notag\\
    &+P(u_i=1|\mathcal{H}_h,i=H)P(i=H)\notag\\
    =&\alpha p\pi_{0,h}+(1-\alpha p)\pi_{1,h}
\end{align}
    and $\widetilde{\pi}_{0,h}=P(u_i=0|\mathcal{H}_h)=1-\widetilde{\pi}_{1,h}$, respectively, for $h=0,1$. From Assumption 1, we have $\pi_{1,0}=\pi_{0,1}\approx 0$ and $\pi_{0,0}=\pi_{1,1}\approx 1$. Thus, we have $\widetilde{\pi}_{1,0}=\widetilde{\pi}_{0,1}\approx \alpha p$ and $\widetilde{\pi}_{1,1}=\widetilde{\pi}_{0,0}\approx 1-\alpha p$.

The fusion rule for the distributed system when all of the sensor decisions are used is given by
\begin{equation}\label{eq:unordered_FR}
    \sum_{i=1}^Nu_i\overset{\mathcal{H}_1}{\underset{\mathcal{H}_0}{\gtrless}}T,
\end{equation}
which follows \cite{niu2005distributed}.
Using the decision rule in \eqref{eq:unordered_FR}, the detection performance can be evaluated in terms of the probability of detection $P_{d,EEOT}^{fc}$ and the probability of false alarm $P_{f,EEOT}^{fc}$ of the FC given below as
\begin{equation}\label{eq:pd-2}
    P_{d,EEOT}^{FC}=\sum_{i=T+1}^N{N \choose i}\pi_{1,1}^i\pi_{0,1}^{N-i}
\end{equation}
and
\begin{equation}\label{eq:pf-2}
    P_{f,EEOT}^{FC}=\sum_{i=T+1}^N{N \choose i}\pi_{1,0}^i\pi_{0,0}^{N-i},
\end{equation}
respectively. Next, we aim at finding the value of optimal $T$ used by the FC in \eqref{eq:unordered_FR} which minimizes the probability of error of both the unordered system and the EEOT-based system. Let $Z=\sum_{i=1}^Nu_i$ denote the number of local decisions that decided $1$. Note that $Z\geq0$. The optimal decision rule at the FC, which is $\frac{\prod_{i=1}^NP(u_i|\mathcal{H}_1)}{\prod_{i=1}^NP(u_i|\mathcal{H}_0)}\overset{\mathcal{H}_1}{\underset{\mathcal{H}_0}{\gtrless}}\frac{\pi_0}{\pi_1}$, can be rewritten as
\begin{align}\label{eq:optimal}
    \left(\frac{\widetilde{\pi}_{1,1}}{\widetilde{\pi}_{1,0}}\right)^Z\left(\frac{1-\widetilde{\pi}_{1,1}}{1-\widetilde{\pi}_{1,0}}\right)^{N-Z}\overset{\mathcal{H}_1}{\underset{\mathcal{H}_0}{\gtrless}}\frac{\pi_0}{\pi_1}.
\end{align}
We make the reasonable assumption that the probability of a sensor being malicious is less than 0.5, i.e., $\alpha<0.5$, and $0\leq p\leq1$ which implies that $\alpha p<0.5$. This implies that $\widetilde{\pi}_{1,1}>\widetilde{\pi}_{1,0}$ (and $\widetilde{\pi}_{0,0}>\widetilde{\pi}_{0,1}$).

Taking the logarithm of both sides of \eqref{eq:optimal}, the optimal decision rule can be rewritten as
\begin{align}\label{eq:optimal_T}
    Z\overset{\mathcal{H}_1}{\underset{\mathcal{H}_0}{\gtrless}}\left[\log\left(\frac{\pi_0}{\pi_1}\right)+N\log\left(\frac{1-\widetilde{\pi}_{1,0}}{1-\widetilde{\pi}_{1,1}}\right)\right]/\log\left(\frac{\widetilde{\pi}_{1,1}(1-\widetilde{\pi}_{1,0})}{\widetilde{\pi}_{1,0}(1-\widetilde{\pi}_{1,1})}\right).
\end{align}
 %Since $\widetilde{\pi}_{1,1}>\widetilde{\pi}_{1,0}$, the left hand side of \eqref{eq:optimal} is an increasing function of $Z$.
Therefore, the optimal threshold $T^*$ at the FC is equal to the right hand side of \eqref{eq:optimal_T},  i.e., $T^*=\left[\log\left(\frac{\pi_0}{\pi_1}\right)+N\log\left(\frac{1-\widetilde{\pi}_{1,0}}{1-\widetilde{\pi}_{1,1}}\right)\right]/\log\left(\frac{\widetilde{\pi}_{1,1}(1-\widetilde{\pi}_{1,0})}{\widetilde{\pi}_{1,0}(1-\widetilde{\pi}_{1,1})}\right)$.

\subsection{Average Number of Transmissions Required for Energy-efficient OT-based System under Attack}
Next, we consider the effect of malicious attacks on the average number of transmissions required by the EEOT scheme. In order to simplify the computation, we find the upper bound (UB) of the average number of transmissions required by finding the lower bound (LB) of the average number of transmissions saved. When the system is under attack, we derive the LB for the average number of transmissions saved in the EEOT-based system. 
%Let $\bar{N}_{s,EE}$ denote the average number of transmissions saved in EEOT-based scheme given as
We first consider the case when the FC decides $\mathcal{H}_1$. It has been derived in \cite{sriranga2018energy} that the average number of transmissions saved to make a final decision is lower bounded by $N/2$ in the absence of the data falsification attacks. Here, we investigate the effect that the attacks have on the lower bound of the expected number of transmissions saved and finding a lower bound for the system under data falsification attacks. It is also shown in this paper that the lower bound we obtain is tight.

We define $k_L^*$ as the minimum number of transmissions required to decide $\mathcal{H}_1$ in the presence of data falsification attacks and it is given in \eqref{eq:k_L_def}. 
Without any loss of generality, let $\lceil T \rceil$ denote the rounding up of $T$ to the closest integer that is greater than or equal to $T$. The average number of transmissions saved when the FC decides $\mathcal{H}_1$ is given as
\begin{small}
\begin{subequations}\label{eq:expected_N_s_EE_L}
\begin{align}
    \bar{N}_{s,1}(\beta)=&E(N-k_L^*)=\sum^{N}_{k=1}(N-k)Pr(k_L^*=k)\\
    =&\sum^{\lceil T \rceil+\beta}_{k=1}(N-k)Pr(k_L^*=k)\notag\\
    &+\sum_{k=\lceil T \rceil+\beta+1}^{N}(N-k)Pr(k_L^*=k)\label{eq:expected_N_s_EE_L_11}\\
    \geq&\sum^{\lceil T \rceil+\beta}_{k=1}(N-k)Pr(k_L^*=k)\label{eq:expected_N_s_EE_L_12}
\\    
\geq&(N-\lceil T \rceil-\beta)Pr(k_L^*\leq \lceil T \rceil+\beta)\\
    =&(N-\lceil T \rceil-\beta)[Pr(\sum_{k=1}^{\lceil T \rceil+\beta}u_{[k]}\geq T|\Gamma_1\geq T)Pr(\Gamma_1\geq T)\notag\\
    &+Pr(\sum_{k=1}^{\lceil T \rceil+\beta}u_{[k]}\geq T|\Gamma_1\leq T)Pr(\Gamma_1\leq T)]\\
    \geq&(N-\lceil T \rceil-\beta)\sum_{h=0,1}Pr(\sum_{k=1}^{\lceil T \rceil+\beta}u_{[k]}\geq T|\Gamma_1\geq T,\mathcal{H}_h)\notag\\
    &\times Pr(\Gamma_1\geq T|\mathcal{H}_h)Pr(\mathcal{H}_h)\\
    =&(N-\lceil T \rceil-\beta)Pr(\sum_{k=1}^{\lceil T \rceil+\beta}u_{[k]}\geq T|\Gamma_1\geq T,\mathcal{H}_1)\notag\\
    &\times Pr(\Gamma_1\geq T|\mathcal{H}_1)Pr(\mathcal{H}_1)\\
    =&(N-\lceil T \rceil-\beta)Pr(\sum_{k=1}^{\lceil T \rceil+\beta}u_{[k]}\geq T|\Gamma_1\geq T,\mathcal{H}_1)Pr(\mathcal{H}_1)\label{eq:expected_N_s_EE_L_f1}\\
    \overset{\bigtriangleup}{=}&f_1(\beta),
\end{align}
\end{subequations}
\end{small}
where $\Gamma_1=\sum^{\lceil T \rceil+\beta}_{i=1}v_{[i]}$ and $Pr(\sum_{k=1}^{\lceil T \rceil+\beta}u_{[k]}\geq T|\Gamma_1\geq T,\mathcal{H}_1)$ can be expressed as
\begin{align}\label{eq:saving_H1_part3}
    Pr(\sum_{k=1}^{\lceil T\rceil+\beta}u_{[i]}\geq T|\Gamma_1\geq T,\mathcal{H}_1)=\sum_{i=0}^{\beta}{\lceil T\rceil+\beta \choose i}\widetilde{\pi}_{0,1}^i\widetilde{\pi}_{1,1}^{\lceil T\rceil+\beta-i}.
\end{align}
Substituting \eqref{eq:saving_H1_part3} in \eqref{eq:expected_N_s_EE_L_f1}, we are able to obtain the LB of the average number of transmissions saved in the network when the FC decides $\mathcal{H}_1$. In going from \eqref{eq:expected_N_s_EE_L_11} to \eqref{eq:expected_N_s_EE_L_12}, the second summation term, which is positive, is dropped. As the difference between the actual average number of transmissions saved and its LB is dependent on the number of terms in the dropped second summation term in \eqref{eq:expected_N_s_EE_L_11}, an appropriate number of terms should be chosen in order to reduce that difference and tighten the LB. Thus, we introduce a variable $\beta$ in \eqref{eq:expected_N_s_EE_L_12} and try to find an appropriate $\beta$ later to prevent the dropped second part of \eqref{eq:expected_N_s_EE_L_11} from being too large so that the LB is tight when the FC decides $\mathcal{H}_1$.
%When $\beta=0$, we can derive the same LB obtained in \cite{sriranga2018energy}. However, $\beta=0$ might not be an appropriate value that allows us to get a tight LB in the presence of attacks. Thus, we aim at finding an appropriate $\beta$ by solving an integer-programming problem. %According to the results obtained in \cite{sriranga2018energy}, $\beta=0$ is an appropriate value for a system to get a tight LB of $\bar{N}_{s,1}$ in the absence of attacks. However, $\beta=0$ might not be an appropriate value when attacks present. 
%Thus, $\beta\neq 0$ is the additional number of transmissions required to achieve a tight LB of $\bar{N}_{s,1}$.

Next, we consider the case when the FC decides $\mathcal{H}_0$. Define $k_U^*$ as the minimum number of transmissions required to decide $\mathcal{H}_0$ and it is given in \eqref{eq:k_U_def}. Let $\lfloor T\rfloor$ denote the rounding down of $T$ to the next lowest integer. Similarly, the average number of transmissions saved when the FC decides $\mathcal{H}_0$ is given as
\begin{small}
\begin{subequations}\label{eq:expected_N_s_EE_L_2}
\begin{align}
    \bar{N}_{s,2}(\beta)=&E(N-k_U^*)=\sum^{N}_{k=1}(N-k)Pr(k_U^*=k)\\
    =&\sum^{N-\lceil T \rceil+\beta}_{k=1}(N-k)Pr(k_U^*=k)\notag\\
    &+\sum_{k=N-\lceil T \rceil+\beta+1}^{N}(N-k)Pr(k_U^*=k)\notag\\
    \geq&\sum^{N-\lceil T \rceil+\beta}_{k=1}(N-k)Pr(k_U^*=k)\\
       \geq&(\lceil T \rceil-\beta)Pr(k_U^*\leq N-\lceil T \rceil+\beta)\\
    =&(\lceil T \rceil-\beta)[Pr(\sum_{k=1}^{N-\lceil T \rceil+\beta}u_{[k]}<\kappa|\Gamma_2>T)Pr(\Gamma_2>T)\notag\\
    &+Pr(\sum_{k=1}^{N-\lceil T \rceil+\beta} u_{[k]}<\kappa|\Gamma_2<T)Pr(\Gamma_2<T)]\\
    \geq&(\lceil T \rceil-\beta)\sum_{h=0,1}Pr(\sum_{k=1}^{N-\lceil T \rceil+\beta} u_{[k]}<\kappa|\Gamma_2<T)\notag\\
    &\times Pr(\Gamma_2<T)\\
    =&(\lceil T \rceil-\beta)Pr(\sum_{k=1}^{N-\lceil T \rceil+\beta} u_{[k]}<\kappa|\Gamma_2<T,\mathcal{H}_0)\notag\\
    &\times Pr(\Gamma_2<T|\mathcal{H}_0)Pr(\mathcal{H}_0)\\
    =&(\lceil T \rceil-\beta)Pr(\sum_{k=1}^{N-\lceil T \rceil+\beta} u_{[k]}<\kappa|\Gamma_2<T,\mathcal{H}_0)Pr(\mathcal{H}_0)\label{eq:expected_N_s_EE_L_2_f2}\\
    \overset{\bigtriangleup}{=}&f_2(\beta),
\end{align}
\end{subequations}
\end{small}
where $\Gamma_2=\sum^{N-\lceil T \rceil+\beta}_{i=1}v_{[i]}$, $\kappa=T-(\lceil T \rceil-\beta)$ and $Pr(\sum_{k=1}^{N-\lceil T \rceil+\beta}u_{[k]}<\kappa|\Gamma_2<T,\mathcal{H}_0)$ can be expressed as
\begin{small}
\begin{align}\label{eq:saving_H0_part3}
    &Pr(\sum_{k=1}^{N-\lceil T \rceil+\beta}u_{[k]}<\kappa|\Gamma_2<T,\mathcal{H}_0)\notag\\
    =&\sum_{i=0}^{\lfloor T \rfloor-\lceil T \rceil+\beta}{N-\lceil T \rceil+\beta \choose i}\widetilde{\pi}_{1,0}^i\widetilde{\pi}_{0,0}^{N-\lceil T \rceil+\beta-i}.
\end{align}
\end{small}

Substituting \eqref{eq:saving_H0_part3} in \eqref{eq:expected_N_s_EE_L_2_f2}, we are able to obtain the LB of the average number of transmissions saved in the network when the FC decides $\mathcal{H}_0$. In a manner similar to the one employed earlier, variable $\beta$ is introduced to ensure that the LB of the average number of transmissions saved is tight when the FC decides $\mathcal{H}_0$. Since only one of the two hypotheses $\mathcal{H}_1$ and $\mathcal{H}_0$ can occur at any given time, the events $k_L^*=k$ and $k_U^*=k$ given hypothesis $\mathcal{H}_1$ or $\mathcal{H}_0$ are disjoint. Hence the total average number of transmissions saved is $N_{s,EEOT}(\beta)=\sum^{N}_{k=1}(N-k)\sum^{1}_{h=0}[Pr(k_U^*=k|\mathcal{H}_h)+Pr(k_L^*=k|\mathcal{H}_h)]Pr(\mathcal{H}_h)=\sum^{N}_{k=1}(N-k)[Pr(k_U^*=k)+Pr(k_L^*=k)]$ and the LB of the average number of transmissions saved is $N_{s,EEOT}^L(\beta)=f_1(\beta)+f_2(\beta)$. When $\beta=0$, the LB derived here reduces to the LB obtained in \cite{sriranga2018energy}. However, $\beta=0$ might not be an appropriate value that allows us to get a tight LB in the presence of attacks. Thus, we aim at finding an optimal $\beta$ so that $N_{s,EEOT}^L(\beta)$ is maximized and the LB becomes tighter. Upon solving the optimization problem given in \eqref{optimization_problem}, we are able to find the optimal $\beta^*$. We denote the set of integers by Z and cast the optimization problem as:
\begin{subequations}\label{optimization_problem}
\begin{align}
& \underset{\bold {\beta}}{\text{max}}
& & f_1(\beta)+f_2(\beta)\\
& \text{s.t.} & & 0\leq \beta\leq min(N-\lceil T \rceil,\lceil T \rceil)\label{constraint1}\\
&&&\beta\in\mathbb{Z}\label{constraint2},
\end{align}
\end{subequations}
The constraint in \eqref{constraint1} is due to the fact that the value of $\beta$ must satisfy both \eqref{beta_range1} and \eqref{beta_range2}, which are derived from \eqref{eq:expected_N_s_EE_L_f1} and \eqref{eq:expected_N_s_EE_L_2_f2}, respectively: 
\begin{align}
    \lceil T \rceil+\beta&\leq N\label{beta_range1}\\
    N-\lceil T \rceil+\beta&\leq N\label{beta_range2}
\end{align}
This is due to the fact that the upper index of the summations in \eqref{eq:expected_N_s_EE_L_f1} and \eqref{eq:expected_N_s_EE_L_2_f2} should be less or equal to $N$.
As the optimization problem in \eqref{optimization_problem} is an integer programming (IP) problem, it is a non-convex optimization problem. However, we have the following lemma which helps us obtain the optimal solution to the optimization problem in \eqref{optimization_problem}.

\begin{theorem}\label{theorem:N_beta}
$N_{s,EEOT}^L(\beta)$ as a function of $\beta$ satisfies either 
\begin{enumerate}
    \item $N_{s,EEOT}^L(\beta)$ is a non-increasing function, $\forall{\beta}\in[0,\min(N-\lceil T \rceil,\lceil T \rceil)]$.
    
    or
    \item There exists a $\beta_l\in\mathbb{Z}$ such that $N_{s,EEOT}^L(\beta)$ is an increasing function $\forall{\beta}\in[0,\beta_l-1]$ and a non-increasing function $\forall{\beta}\in[\beta_l,\min(N-\lceil T \rceil,\lceil T \rceil)]$.
\end{enumerate}
\end{theorem}
\begin{IEEEproof}
 Let $g_1(\beta)=\sum_{i=0}^{\beta}{\lceil T\rceil+\beta \choose i}\widetilde{\pi}_{0,1}^i\widetilde{\pi}_{1,1}^{\lceil T\rceil+\beta-i}$ and $g_2(\beta)=\sum_{i=0}^{\lfloor T \rfloor-\lceil T \rceil+\beta}{N-\lceil T \rceil+\beta \choose i}\widetilde{\pi}_{1,0}^i\widetilde{\pi}_{0,0}^{N-\lceil T \rceil+\beta-i}$. Hence, we have
 \begin{align}
     g_1(\beta+1)=\sum_{i=0}^{\beta+1}{\lceil T\rceil+\beta+1 \choose i}\widetilde{\pi}_{0,1}^i\widetilde{\pi}_{1,1}^{\lceil T\rceil+\beta+1-i}
 \end{align}
 \begin{align}
     g_2(\beta+1)=\sum_{i=0}^{T_d+\beta+1}{N-\lceil T \rceil+\beta+1 \choose i}\widetilde{\pi}_{1,0}^i\widetilde{\pi}_{0,0}^{N-\lceil T \rceil+\beta+1-i},
 \end{align}
 where $T_d=\lfloor T \rfloor-\lceil T \rceil=-1$, $\widetilde{\pi}_{0,1}=\widetilde{\pi}_{1,0}=\alpha p$ and $\widetilde{\pi}_{0,0}=\widetilde{\pi}_{1,1}=1-\alpha p$ based on Assumption 1. $g_1(\beta+1)$ and $g_2(\beta+1)$ can be expressed in terms of $g_1(\beta)$ and $g_2(\beta)$ that are respectively given as
 \begin{align}\label{eq:g1}
     g_1(\beta+1)=g_1(\beta)+A(\beta)
 \end{align}
 and
 \begin{align}\label{eq:g2}
     g_2(\beta+1)=g_2(\beta)+B(\beta),
 \end{align}
 where $A(\beta)={\lceil T \rceil+\beta \choose \beta+1}(\alpha p)^{\beta+1}(1-\alpha p)^{\lceil T \rceil}$ and $B(\beta)={N-\lceil T \rceil+\beta \choose \beta+T_d+1}(\alpha p)^{\beta+T_d+1}(1-\alpha p)^{N-\lceil T \rceil-T_d}$. The derivations of \eqref{eq:g1} and \eqref{eq:g2} are relegated to Appendix~\ref{Appendix:additioanl}. It is evident that if
 \begin{align}\label{eq:decreasing_condition1}
     [f_1(\beta+1)+f_2(\beta+1)]-[f_1(\beta)+f_2(\beta)]< 0,
 \end{align}
 then $N_{s,EEOT}^L(\beta)> N_{s,EEOT}^L(\beta+1)$. By rewriting \eqref{eq:decreasing_condition1} using \eqref{eq:g1} and \eqref{eq:g2}, we obtain the inequality 
  \begin{align}\label{eq:decreasing_condition}
     D(\beta)> D_2(\beta),
 \end{align}
 where $D(\beta)=\pi_1g_1(\beta)+\pi_0g_2(\beta)$, $D_2(\beta)=\pi_1h_{1}(\beta)+\pi_0h_{2}(\beta)$, $\pi_1=Pr(\mathcal{H}_1)$, $\pi_0=Pr(\mathcal{H}_0)$, $h_{1}(\beta)=(N-\lceil T\rceil-\beta-1)A(\beta)$ and $h_{2}(\beta)=(\lceil T\rceil-\beta-1)B(\beta)$.

We proceed to show that if $D(\beta)> D_2(\beta)$ is true, then $D(\beta+1)> D_2(\beta+1)$ is also true. Using the expressions in \eqref{eq:g1}, \eqref{eq:g2} and \eqref{eq:decreasing_condition}, we rewrite $D(\beta+1)> D_2(\beta+1)$ as
\begin{align}\label{eq:D_beta_D2_beta}
    D(\beta)+\pi_1A(\beta)+\pi_0B(\beta)>D_2(\beta)-\pi_1A(\beta)-\pi_0B(\beta).
\end{align}
By reformulating \eqref{eq:D_beta_D2_beta}, we have
\begin{align}\label{eq:beta+1}
   2[\pi_1A(\beta)+\pi_0B(\beta)]>D_2(\beta)-D(\beta).
\end{align}
Due to the assumption that $D(\beta)> D_2(\beta)$, $A(\beta)\geq0$ and $B(\beta)\geq0$, the left hand side of \eqref{eq:beta+1} is greater than or equal to 0 and the right hand side of \eqref{eq:beta+1} is smaller than 0. Therefore, \eqref{eq:beta+1} is always true if $D(\beta)> D_2(\beta)$ is true. In other words, if $N_{s,EEOT}^L(\beta)>N_{s,EEOT}^L(\beta+1)$, we always have $N_{s,EEOT}^L(\beta+1)>N_{s,EEOT}^L(\beta+2)$. 

Let $\beta=\beta_s$ be the smallest $\beta$ for which $N_{s,EEOT}^L(\beta)>N_{s,EEOT}^L(\beta+1)$. If $\beta_s=0$, $N_{s,EEOT}^L(\beta)$ is a decreasing function of $\beta$. If $\beta_s>0$, $N_{s,EEOT}^L(\beta)$ is a non-decreasing function of $\beta$ when $\beta\in[0,\beta_s-1]$ and a decreasing function of $\beta$ when $\beta\in[\beta_s,\min(N-\lceil T \rceil,\lceil T \rceil)]$. Let $\beta=\beta_l$ be the largest $\beta$ for which $N_{s,EEOT}^L(\beta)<N_{s,EEOT}^L(\beta+1)$. The above statement is equivalent to that made in Theorem~\ref{theorem:N_beta} about the monotonicity of $N_{s,EEOT}(\beta)$. This completes our proof.
\end{IEEEproof}

According to Theorem~\ref{theorem:N_beta}, the optimal solution $\beta^*$ to the optimization problem in \eqref{optimization_problem} is the smallest $\beta$ for which the inequality $D(\beta)\geq D_2(\beta)$ holds, and the LB of the number of transmissions saved is then given as
\begin{align}\label{eq:optimal_LB}
     N_{s,EEOT}^L=f_1(\beta^*)+f_2(\beta^*).
\end{align}
Therefore, the tight UB of the average number of transmissions required is $N_{t,EEOT}^U=N-N_{s,EEOT}^L$.

% \begin{figure}[htbp]
% \centerline{\includegraphics[width=\linewidth,height=12em]{3D_plot.PNG}}
% \caption{$N_{s,EEOT}/N$ as a function of $p$ and $\beta$ when $T=50.5$, $\alpha=0.3$, $N=100$ and $\pi_1=0.5$.}
% \end{figure}

\section{Numerical and Simulation Results}
\label{sec:simulation_EEOT}
\begin{figure}[htbp]
\centerline{\includegraphics[width=\linewidth,height=12em]{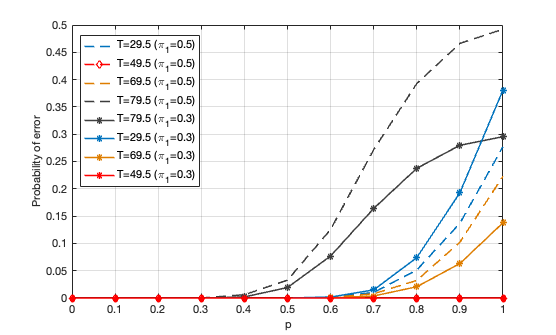}}
\caption{$P_e$ as a function of $p$ with different values of $T$ for the EEOT-based system for $\pi_1=0.3$ and $\pi_1=0.5$.}
\label{fig:Pe_0.3_0.5}
\end{figure}

\begin{figure}[htbp]
\centerline{\includegraphics[width=\linewidth,height=20em]{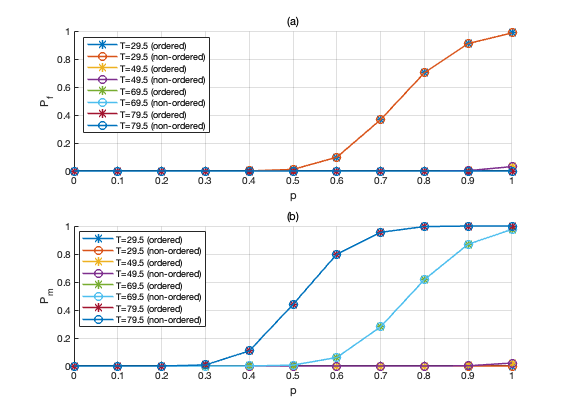}}
\caption{(a) $P_f$ as a function of $p$ for different values of $T$, for the EEOT-based system and the unordered system when $\pi_1=\pi_0=0.5$. (b) $P_m$ as a function of $p$ with different values of $T$ for the EEOT-based system and the unordered system when $\pi_1=\pi_0=0.5$.}
\label{fig:subplot_pf_pm}
\end{figure}

In this section, we present some numerical and simulation results to support our theoretical analysis and show the detection performance, the number of transmissions required and the improved UB compared with the UB obtained in \cite{sriranga2018energy}. We assume that the number of sensors in the network is $N=100$. 
Fig. \ref{fig:Pe_0.3_0.5} shows the probability of error as a function of $p$ in the EEOT-based system. The system with the threshold closest to the optimal threshold $T^*$ ($T^*$ is roughly $N/2$), as compared to other systems, has the lowest error probability, which is in accordance with the conclusion that we obtained about the optimal threshold $T^*$.\footnote{The threshold closest to the optimal threshold $T^*$ is 49.5 in Fig. \ref{fig:subplot_pf_pm} when $\pi_1=0.5$ and $\pi_1=0.3$.} Fig. \ref{fig:subplot_pf_pm} shows the probabilities of false alarm and missed detection for the EEOT-based system and the unordered system. For the same parameter values, we can observe that both EEOT-based and unordered systems have the same probabilities of false alarm and missed detection, thus yielding the same error probability. This is in accordance with Lemma~\ref{lem:detect_perf_ot}.

\begin{figure}[htbp]
\centerline{\includegraphics[width=\linewidth,height=12em]{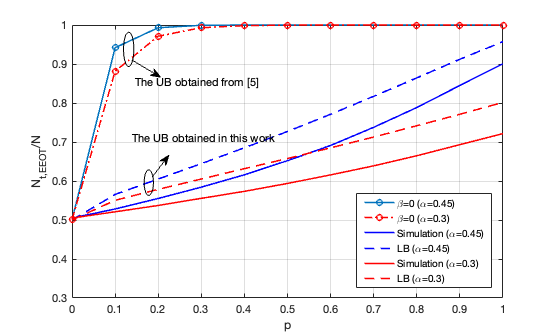}}
\caption{Benchmarking upper bounds for the fraction of the number of transmissions required $N_{t,EEOT}/N$ as a function of $p$ with different values of $\alpha$ when $\pi_1=0.5$.}
\label{fig:tight_lB}
\end{figure}

\begin{figure}[htbp]
\centerline{\includegraphics[width=\linewidth,height=20em]{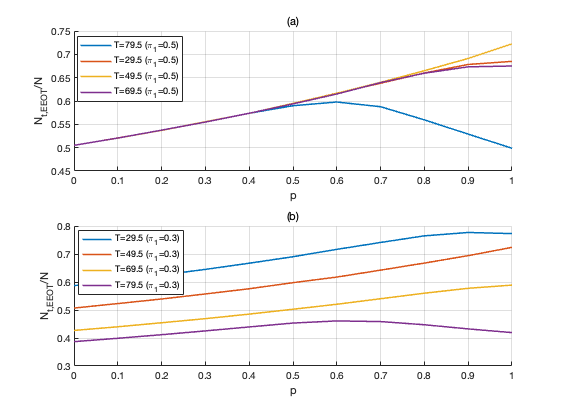}}
\caption{(a) $N_{t,EEOT}/N$ as a function of $p$ with different values of $T$ when $\alpha=0.3$ and $\pi_1=0.3$. (b) $N_{t,EEOT}/N$ as a function of $p$ with different values of $T$ when $\alpha=0.3$ and $\pi_1=0.5$.}
\label{fig:Fraction_saved_differient_T_pi_0.3_0.5}
\end{figure}

Fig. \ref{fig:tight_lB} shows that the UB we obtained is a relatively tight UB compared with the UB obtained in \cite{sriranga2018energy} for the average fraction of number of transmissions required as a function of the attacking probability $p$ in the EEOT-based system. Fig. \ref{fig:Fraction_saved_differient_T_pi_0.3_0.5} presents the average fraction of transmissions required $N_{t,EEOT}/N$ in the EEOT-based system as a function of $p$ for different values of prior probability and $T$ when $\alpha=0.3$. We observe from Fig. \ref{fig:Fraction_saved_differient_T_pi_0.3_0.5} (a) that when $T\rightarrow T^*$ ($T^*$ here is roughly $N/2$), the system is most likely to have the highest transmissions required in the network if the prior probabilities of both hypotheses are $0.5$. However, when the prior probabilities change, the value of $T$ that results in the highest transmissions required might also change. It is clear that a smaller $T$ results in a larger number of transmissions required to decide $\mathcal{H}_0$ and a smaller number of transmissions required to decide $\mathcal{H}_1$. For a relatively small $\pi_1$ that satisfies $\pi_1<0.5$, the probability of the FC deciding $\mathcal{H}_0$ is higher. Consequently, as shown in Fig. \ref{fig:Fraction_saved_differient_T_pi_0.3_0.5} (b), the system that uses $T=29.5$ has higher transmissions required when compared to the system that uses $T=49.5$ given $\pi_1=0.3$. Thus, there is a relationship between the average number of transmissions needed and the detection performance of the system. With an appropriately designed threshold used by the FC, it is possible to save transmissions while still guaranteeing the quality of the decision.

% \begin{figure}[htbp]
% \centerline{\includegraphics[width=\linewidth,height=12em]{Fraction_saved_differient_T_pi_0.3_0.5.png}}
% \caption{$N_{s,EEOT}/N$ as a function of $p$ with different values of $T$ when $\alpha=0.3$ and $\pi_1=0.3,0.5$.}
% \label{fig:Fraction_saved_differient_T_pi_0.3_0.5}
% \end{figure}

% \begin{figure}[htbp]
% \centerline{\includegraphics[width=\linewidth,height=15em]{ordered_non_ordered_comparsion_pi_0.5.png}}
% \caption{$P_e$ as a function of $p$ with different values of $T$ for EEOT-based system and non-ordered based system when $\pi_1=\pi_0=0.5$ and $N=100$.}
% \end{figure}

\section{Conclusion}
\label{sec:conclusion_EEOT}
In this work, we considered the data falsification attack problem in an EEOT-based distributed detection system. We evaluated the performance of the EEOT-based system via the detection performance and the maximum number of transmissions required in the presence of data falsification attacks. We showed that the detection performance of a system using ordered transmissions is unaffected in the presence of data falsification attacks and the probability of error of the EEOT-based system was derived. We also found a tight UB on the average number of transmissions required under attack. Moreover, the optimal threshold for the FC was obtained in this work. Numerical and simulation results indicate that it is possible to reduce transmissions while still ensuring the quality of the decision with an appropriately designed threshold.

\appendices
\section{Proof of Lemma~\ref{lem:detect_perf_ot}}
\label{Appendix:EEOT}
Recall that $k_U^*$ and $k_L^*$ are the minimum number of transmissions required to decide $\mathcal{H}_0$ and $\mathcal{H}_1$, respectively. These are given as follows.

\begin{equation}
k_U^*=\min\limits_{1\leq k\leq N}\left\{\sum_{i=1}^ku_{[i]}< T -(N-k)\right\}\label{eq:k_U_def}
\end{equation} 
\begin{equation}k_L^*=\min\limits_{1\leq k\leq N}\left\{\sum_{i=1}^ku_{[i]}\geq T\right\}.\label{eq:k_L_def}
\end{equation}
It is obvious that $k_U^*$ and $k_L^*$ can not exist at the same time. If $k_U^*$ is valid, the FC decides hypotheses $\mathcal{H}_0$, and if $k_L^*$ is valid, the FC decides hypotheses $\mathcal{H}_1$. Since only one of the two hypotheses $\mathcal{H}_1$ and $\mathcal{H}_0$ can occur at any given time, only one of $k_L^*$ or $k_U^*$ is valid at any given time. Let $Z_U$, $Z_L$ denote the upper bound and lower bound of $Z=\sum_{i=1}^Nu_{[i]}$, respectively, if $k_U^*$ or $k_L^*$ is valid. 
Due to the fact that $u_{[i]}\in\{0,1\}$ for $\forall i\in\{1,2,\dots,N\}$, we have 
\begin{equation}\label{eq:Z-U}
    Z_U=\sum_{i=1}^{k_U^*}u_{[i]}+(N-k_U^*)\geq \sum_{i=1}^Nu_{[i]}=Z
\end{equation}
if $k_U^*$ is valid and
\begin{equation}\label{eq:Z-l}
    Z_L=\sum_{i=1}^{k_L^*}u_{[i]}\leq \sum_{i=1}^Nu_{[i]}=Z
\end{equation}
if $k_L^*$ is valid. According to the fusion rule given in \eqref{eq:EE-rule}, the FC decides hypothesis $\mathcal{H}_0$ if $Z_U< T$, and hypothesis $\mathcal{H}_1$ if $Z_L\geq T$. Since only one of $k_U^*$ and $k_L^*$ is valid, we have $Pr(k_U^* \text{ is valid}) + Pr(k_L^* \text{ is valid}) = 1$, which is equivalent to $Pr(Z_U< T) + Pr(Z_L\geq T) = 1$.

If the FC decides $\mathcal{H}_1$ for an unordered system, we have
\begin{subequations}
\begin{align}
    Z\geq T\overset{implies}{\Longrightarrow}&\sum^{k}_{i=1}u_{[i]}+\sum^{N}_{i=k+1}u_{[i]}\geq T\\
    \Longrightarrow&\sum^{k}_{i=1}u_{[i]}\geq T-\sum^{N}_{i=k+1}u_{[i]}\geq T-(N-k)\\
    \Longrightarrow&\sum^{k}_{i=1}u_{[i]}\geq T-(N-k)\label{eq:z>lamba}
\end{align}
\end{subequations}
for $\forall k$. We could observe from \eqref{eq:k_U_def} that  $k^*_U$ is not valid when $Z\geq T$. So if $Z \geq T$, we have $Pr(Z_U< T) = 0$ and $Pr(Z_L\geq T) = 1$.
Hence, we can conclude that $Pr(Z_L\geq T|Z\geq T,\mathcal{H}_j)=1$. Upon following a similar analysis, we obtain $Pr(Z\geq T|Z_L\geq T,\mathcal{H}_j)=1$. This allows us to calculate $Pr(Z_L\geq T|\mathcal{H}_j)$ according to Bayes' rule which is given as
% \begin{equation}
\begin{align}
        Pr(Z_L\geq T|\mathcal{H}_j)&=\frac{Pr(Z_L\geq T|Z\geq T,\mathcal{H}_j)Pr(Z\geq T|\mathcal{H}_j)}{Pr(Z\geq T|Z_L\geq T,\mathcal{H}_j)} \nonumber \\
    &=Pr(Z\geq T|\mathcal{H}_j).
\end{align}
% \end{equation}
Similarly, we obtain $Pr(Z_U< T|\mathcal{H}_j)=Pr(Z< T|\mathcal{H}_j)$. Hence, the probability of error of the EEOT-based system is given as
% \begin{subequations}
\begin{align}
    P_e^{(OT)}&=\pi_0Pr(Z_L\geq T|\mathcal{H}_0)+\pi_1Pr(Z_U< T|\mathcal{H}_1) \nonumber \\
    &=\pi_0Pr(Z\geq T|\mathcal{H}_0)+\pi_1Pr(Z< T|\mathcal{H}_1)=P_e^{(opt)},
\end{align}
% \end{subequations}
where $P_e^{(opt)}$ is the probability of error of the unordered system.

\section{}
\label{Appendix:additioanl}
Since $g_1(\beta)=\sum_{i=0}^{\beta}{\lceil T\rceil+\beta \choose i}(\alpha p)^i(1-\alpha p)^{\lceil T\rceil+\beta-i}$, we have
\begin{subequations}
 \begin{align}
     g_1(\beta+1)=&\sum_{i=0}^{\beta+1}{\lceil T\rceil+\beta+1 \choose i}(\alpha p)^i(1-\alpha p)^{\lceil T\rceil+\beta+1-i}\\
     =&(1-\alpha p)^{\lceil T\rceil+\beta+1}+\sum_{i=1}^{\beta+1}\left[{\lceil T\rceil+\beta \choose i}+{\lceil T\rceil+\beta \choose i-1}\right]\notag\\
     &\times(\alpha p)^i(1-\alpha p)^{\lceil T\rceil+\beta+1-i}\\
     =&\sum_{i=0}^{\beta+1}{\lceil T\rceil+\beta \choose i}(\alpha p)^i(1-\alpha p)^{\lceil T\rceil+\beta+1-i}\notag\\
     &+\sum_{i=1}^{\beta+1}{\lceil T\rceil+\beta \choose i-1}(\alpha p)^i(1-\alpha p)^{\lceil T\rceil+\beta+1-i}\\
     =&{\lceil T\rceil+\beta \choose \beta+1}(\alpha p)^{\beta+1}(1-\alpha p)^{\lceil T\rceil}+(1-\alpha p)g_1(\beta)\notag\\
     &+\sum_{i=0}^{\beta}{\lceil T\rceil+\beta \choose i}(\alpha p)^i(1-\alpha p)^{\lceil T\rceil+\beta-i}(\alpha p)\\
     =&g_1(\beta)+{\lceil T\rceil+\beta \choose \beta+1}(\alpha p)^{\beta+1}(1-\alpha p)^{\lceil T\rceil}
\end{align}
based on Pascal's rule. Following a similar sequence of steps, we can obtain
\begin{align}
    g_2(\beta+1)=g_2(\beta){N-\lceil T \rceil+\beta \choose \beta+T_d+1}(\alpha p)^{\beta+T_d+1}(1-\alpha p)^{N-\lceil T \rceil-T_d}.
\end{align}
\end{subequations}

\bibliography{refer.bib}

\begin{thebibliography}{10}
\providecommand{\url}[1]{#1}
\csname url@samestyle\endcsname
\providecommand{\newblock}{\relax}
\providecommand{\bibinfo}[2]{#2}
\providecommand{\BIBentrySTDinterwordspacing}{\spaceskip=0pt\relax}
\providecommand{\BIBentryALTinterwordstretchfactor}{4}
\providecommand{\BIBentryALTinterwordspacing}{\spaceskip=\fontdimen2\font plus
\BIBentryALTinterwordstretchfactor\fontdimen3\font minus
  \fontdimen4\font\relax}
\providecommand{\BIBforeignlanguage}[2]{{%
\expandafter\ifx\csname l@#1\endcsname\relax
\typeout{** WARNING: IEEEtran.bst: No hyphenation pattern has been}%
\typeout{** loaded for the language `#1'. Using the pattern for}%
\typeout{** the default language instead.}%
\else
\language=\csname l@#1\endcsname
\fi
#2}}
\providecommand{\BIBdecl}{\relax}
\BIBdecl

\bibitem{rago1996censoring}
C.~Rago, P.~Willett, and Y.~Bar-Shalom, ``Censoring sensors: A
  low-communication-rate scheme for distributed detection,'' \emph{IEEE
  Transactions on Aerospace and Electronic Systems}, vol.~32, no.~2, pp.
  554--568, 1996.

\bibitem{bandyopadhyay2003energy}
S.~Bandyopadhyay and E.~J. Coyle, ``An energy efficient hierarchical clustering
  algorithm for wireless sensor networks,'' in \emph{IEEE INFOCOM 2003.
  Twenty-second Annual Joint Conference of the IEEE Computer and Communications
  Societies (IEEE Cat. No. 03CH37428)}, vol.~3.\hskip 1em plus 0.5em minus
  0.4em\relax IEEE, 2003, pp. 1713--1723.

\bibitem{bajovic2011sensor}
D.~Bajovic, B.~Sinopoli, and J.~Xavier, ``Sensor selection for event detection
  in wireless sensor networks,'' \emph{IEEE Transactions on Signal Processing},
  vol.~59, no.~10, pp. 4938--4953, 2011.

\bibitem{blum2008energy}
R.~S. Blum and B.~M. Sadler, ``Energy efficient signal detection in sensor
  networks using ordered transmissions,'' \emph{IEEE Transactions on Signal
  Processing}, vol.~56, no.~7, pp. 3229--3235, 2008.

\bibitem{sriranga2018energy}
N.~Sriranga, K.~G. Nagananda, R.~S. Blum, A.~Saucan, and P.~K. Varshney,
  ``Energy-efficient decision fusion for distributed detection in wireless
  sensor networks,'' in \emph{2018 21st International conference on information
  fusion (FUSION)}.\hskip 1em plus 0.5em minus 0.4em\relax IEEE, 2018, pp.
  1541--1547.

\bibitem{braca2011single}
P.~Braca, S.~Marano, and V.~Matta, ``Single-transmission distributed detection
  via order statistics,'' \emph{IEEE Transactions on Signal Processing},
  vol.~60, no.~4, pp. 2042--2048, 2011.

\bibitem{gupta2020ordered2}
S.~S. Gupta and N.~B. Mehta, ``Ordered transmissions schemes for detection in
  spatially correlated wireless sensor networks,'' \emph{IEEE Transactions on
  Communications}, vol.~69, no.~3, pp. 1565--1577, 2020.

\bibitem{rawas2011energy}
Z.~N. Rawas, Q.~He, and R.~S. Blum, ``Energy-efficient noncoherent signal
  detection for networked sensors using ordered transmissions,'' in \emph{2011
  45th Annual Conference on Information Sciences and Systems}.\hskip 1em plus
  0.5em minus 0.4em\relax IEEE, 2011, pp. 1--5.

\bibitem{hesham2012distributed}
L.~Hesham, A.~Sultan, M.~Nafie, and F.~Digham, ``Distributed spectrum sensing
  with sequential ordered transmissions to a cognitive fusion center,''
  \emph{IEEE Transactions on Signal Processing}, vol.~60, no.~5, pp.
  2524--2538, 2012.

\bibitem{chen2020optimal}
Y.~Chen, R.~S. Blum, and B.~M. Sadler, ``Optimal quickest change detection in
  sensor networks using ordered transmissions,'' in \emph{2020 IEEE 21st
  International Workshop on Signal Processing Advances in Wireless
  Communications (SPAWC)}.\hskip 1em plus 0.5em minus 0.4em\relax IEEE, 2020,
  pp. 1--5.

\bibitem{gupta2020ordered}
S.~S. Gupta, S.~K. Pallapothu, and N.~B. Mehta, ``Ordered transmissions for
  energy-efficient detection in energy harvesting wireless sensor networks,''
  \emph{IEEE Transactions on Communications}, vol.~68, no.~4, pp. 2525--2537,
  2020.

\bibitem{chen2020ordered}
Y.~Chen, B.~M. Sadler, and R.~S. Blum, ``Ordered gradient approach for
  communication-efficient distributed learning,'' in \emph{2020 IEEE 21st
  International Workshop on Signal Processing Advances in Wireless
  Communications (SPAWC)}.\hskip 1em plus 0.5em minus 0.4em\relax IEEE, 2020,
  pp. 1--5.

\bibitem{6582732}
A.~Vempaty, L.~Tong, and P.~K. Varshney, ``Distributed inference with byzantine
  data: State-of-the-art review on data falsification attacks,'' \emph{IEEE
  Signal Processing Magazine}, vol.~30, no.~5, pp. 65--75, 2013.

\bibitem{kailkhura2016data}
B.~Kailkhura, S.~Brahma, and P.~K. Varshney, ``Data falsification attacks on
  consensus-based detection systems,'' \emph{IEEE Transactions on Signal and
  Information Processing over Networks}, vol.~3, no.~1, pp. 145--158, 2016.

\bibitem{quan2022enhanced}
C.~Quan, B.~Geng, Y.~S. Han, and P.~K. Varshney, ``Enhanced audit bit based
  distributed bayesian detection in the presence of strategic attacks,''
  \emph{IEEE Transactions on Signal and Information Processing over Networks},
  vol.~8, pp. 49--62, 2022.

\bibitem{niu2005distributed}
R.~Niu and P.~K. Varshney, ``Distributed detection and fusion in a large
  wireless sensor network of random size,'' \emph{EURASIP Journal on Wireless
  Communications and Networking}, vol. 2005, no.~4, pp. 1--11, 2005.

\end{thebibliography}
\bibliographystyle{IEEEtran}

\end{document}